\documentclass[12pt,letterpaper]{JHEP}
\usepackage{epsfig,cite}

\title{Positive Vacuum Energy and the $N$-bound}

\author{
Raphael Bousso \\
Institute for Theoretical Physics\\
University of California, Santa Barbara, California 93106-4030\\
E-mail: \email{bousso@itp.ucsb.edu}}

\abstract{We argue that the total observable entropy is bounded by the
inverse of the cosmological constant.  This holds for all space-times
with a positive cosmological constant, including cosmologies dominated
by ordinary matter, and recollapsing universes.  The argument involves
intermediate steps which may be of interest in their own right.  We
note that entropy cannot be observed unless it lies both in the past
and in the future of the observer's history.  This truncates
space-time to a diamond-shaped subset well-suited to the application
of the covariant entropy bound.  We further require, and derive, a
novel Bekenstein-like bound on matter entropy in asymptotically
de~Sitter spaces.  Our main result lends support to the proposal that
universes with positive cosmological constant are described by a
fundamental theory with only a finite number of degrees of freedom.}

\preprint{\hepth{0010252} \\ NSF-ITP-00-113}

\begin{document}

\section{Introduction}
\label{sec-intro}

\subsection{Banks's proposal}
\label{sec-banks}

Banks~\cite{Ban00} has proposed that the cosmological constant should
not be viewed as an effective parameter to be derived in a theoretical
framework like QFT or string theory.  Instead, it is determined as the
inverse of the number of degrees of freedom, $N$, in the fundamental
theory.%
\footnote{In this paper, $N$ always denotes the number of degrees of
freedom; it should not be confused with the size of a gauge group, or
the level of supersymmetry.  The space-time dimension is taken to be 4
in order to keep equations simple, but generalization to arbitrary
dimensions is trivial.  Planck units are used throughout.}
It should thus be considered an input parameter at the most
fundamental level of physics.

The proposal can be motivated as follows.  In the presence of a
positive cosmological constant, $\Lambda$, the universe tends to
evolve to empty de~Sitter space.  de~Sitter space has a finite entropy
$S = 3\pi / \Lambda$, given by the area of the cosmological horizon.
Thus the universe is most economically described by a theory with the
corresponding number of degrees of freedom, $N = 3\pi / \Lambda$.
Conversely, a quantum gravity theory with a finite number of degrees
of freedom, $N$, requires for consistency a cosmological constant
$\Lambda = 3\pi/N$ to provide a geometric entropy cutoff.

The $\Lambda$-$N$ correspondence does not solve the cosmological
constant problem except by fiat.  It is not clear why the fundamental
theory should happen to possess the enormous but finite number of
degrees of freedom $N \sim 10^{122}$ that corresponds to the
observationally favoured value of the cosmological constant.  But the
proposal offers a radical, and potentially fruitful, change of
perspective.

Its most profound implication is the following: A quantum gravity
theory with an infinite number of degrees of freedom, such as M
theory, cannot describe space-times with a positive cosmological
constant.%
\footnote{For a quantum theory, $N$ is defined to be the logarithm of
the dimension of Hilbert space; thus, a theory with finite $N$ has a
finite-dimensional Hilbert space.  Superstring theories, and current
non-perturbative proposals for M-theory~\cite{BanFis96,Sus97,Mal97},
have an infinite-dimensional Hilbert space.  (Indeed, even a single
harmonic oscillator has an infinite-dimensional Hilbert space.)}
This is consistent with the fact that no stable de~Sitter vacua are
known in M theory.  If the proposal is correct, this gap would not be
due to our limited understanding of the theory, but must be ascribed
to an obstruction in principle.

The correspondence thus suggests that one should look for a theory
with finite $N$ that is self-consistent and {\em complete\/}; i.e., it
will not do to impose a naive cut-off on an $N=\infty$ theory.  If
such theories exist for arbitrarily large values of $N$, one might
expect them to limit to M theory.  However, finite $N$ theories will
contain certain qualitative features, such as positive vacuum energy
and perhaps supersymmetry breaking,%
\footnote{Ref.~\cite{Ban00} also explores the possibility of a
connection between finite $N$ and supersymmetry (SUSY) breaking,
noting that no stable SUSY-violating vacuum states have been firmly
identified in M theory (see, however,
Refs.~\cite{KacKum98,KacSil98a,KacSil98b}).  One may therefore
speculate that SUSY breaking can only occur in theories with finite
$N$, and that both the SUSY and the vacuum energy scales arise from
finite $N$.  One then faces the challenge of explaining why the SUSY
scale is much larger than $N^{-1}$.  We shall not pursue the
connection with supersymmetry in the present paper.}
which would be entirely absent in the infinite $N$ limit and could not
have been studied there.

How can the proposal be tested?  It asserts that a universe with
$\Lambda>0$ is a system with $N = 3\pi/\Lambda$ degrees of freedom.
Unfortunately, this cannot be verified at the semi-classical level, as
we have no understanding what the true degrees of freedom are.
However, a system with $N$ degrees of freedom certainly cannot have
entropy greater than $N$.  Thus, the $\Lambda$-$N$ correspondence
predicts that a universe with $\Lambda >0$ cannot have entropy greater
than $N = 3\pi/\Lambda$.  We call this prediction the {\em $N$-bound}.
It can be tested.

It is not difficult to see that the $N$-bound is true for vacuum
solutions like de~Sitter space (a trivial case).  Moreover, one can
argue that it is satisfied for all space-times which are
asymptotically de~Sitter at late times, by the generalized second law
of thermodynamics.  Indeed, in Ref.~\cite{Ban00} the $\Lambda$-$N$
correspondence was conjectured to apply only within this class of
space-times.  This includes, for example, the $\Lambda > 0$ flat
Friedman-Robertson-Walker (FRW) solution which appears to describe our
universe: it starts out with a big bang and is initially radiation- or
matter-dominated; then the matter is diluted by the cosmological
expansion; and at some time (as it happens, roughly now) the vacuum
energy---which is not diluted---starts to dominate and leads the
universe to evolve towards empty de~Sitter space in the far future.

However, solutions with $\Lambda > 0$ need not necessarily become
de~Sitter at late times.  Consider, for example, the time reversal of
the cosmological solution just described: it starts out as empty
de~Sitter; then more and more matter condenses, which eventually
causes the space-time to collapse in a big crunch.  This illustrates,
in particular, that one can never be sure to have reached the safety
of asymptotic de~Sitter space; there is always the possibility of a
huge collapsing shell of matter that cannot be seen yet but will cause
an apocalypse in the future.  Another example is a $\Lambda > 0$
closed FRW universe.  Given a sufficiently large matter density, the
cosmological constant will not be strong enough to prevent recollapse.
Indeed, $\Lambda$ might be a negligible contribution to the total
energy density at all times.

These $\Lambda>0$ solutions are perfectly valid from the perspective
of semi-classical gravity.  Many of them are physically quite
reasonable, and we would find it unconvincing to exclude them {\em a
priori}.  In some cases, a small perturbation can make all the
difference between collapse and expansion to asymptotically de Sitter
space.  These arguments lead us to advocate a stronger version of
Banks's proposal.  We conjecture that the $\Lambda$-$N$ correspondence
holds for all $\Lambda>0$ universes, including those that do not
evolve to de~Sitter in the future.  But if de~Sitter is not the `final
state', the second law will be of no help, and it is no longer obvious
that the $N$-bound holds.  The $N$-bound is thus a non-trivial
prediction of the $\Lambda$-$N$ correspondence.

Indeed, at first sight, some solutions may appear to have entropy
greater than $N$, in contradiction with the correspondence.
Nevertheless, it will be argued in this paper that the $N$-bound is
valid for all universes with $\Lambda > 0$.  This statement is far
from obvious, and its proof will be seen to require a number of
non-trivial intermediate results.  Therefore, our conclusion may be
viewed as evidence in favour of the proposed correspondence.

\subsection{Outline}

Our goal is to prove the following conjecture:

\paragraph{$N$-bound}
{\em In any universe with a positive cosmological constant $\Lambda$
(as well as arbitrary additional matter that may well dominate
at all times) the observable entropy $S$ is bounded by}
\begin{equation}
N = 3 \pi / \Lambda.
\end{equation}
Here $S$ includes both matter and horizon entropy, but excludes
entropy that cannot be observed in a causal experiment.  Note that $N$
is the Bekenstein-Hawking entropy of empty de~Sitter space.  The bound
becomes trivial in the limit of vanishing cosmological constant.  As
we have argued above, an independent proof of the $N$-bound provides
strong support to the proposed $\Lambda$-$N$
correspondence~\cite{Ban00}; hence, the correspondence will not be
used anywhere in the paper.

In Sec.~\ref{sec-cd} we ask what constitutes observable entropy.  We
argue that one should restrict attention to the {\em causal diamond\/}
of an observer: the space-time region that can be both influenced and
seen by an observer.  Thus, the observable entropy lies in a region
bounded by the past and future light cones from the endpoints of the
observer's world line.

The covariant entropy bound~\cite{Bou99b,Bou99c,Bou99d,FlaMar99},
reviewed in Sec.~\ref{sec-ceb}, can be applied to the cones bounding
the observable region.  This turns out to imply only $S\leq 2N$,
however, which does not quite suffice.  In Sec.~\ref{sec-dbound}, we
derive a novel bound on the entropy of matter systems in de Sitter
space, the `D-bound', which can be tighter than the covariant bound.
In Sec.~\ref{sec-nbound} we argue that the two bounds can be combined
to imply the $N$-bound.  The results are discussed in
Sec.~\ref{sec-summary}.

Banks's discussion~\cite{Ban00} of the consistency of his proposal
involved many of the considerations that enter our derivation of the
$N$-bound.  The arguments presented here are strongly influenced by
Susskind's emphasis on the operational meaning of physical quantities.
The covariant entropy bound~\cite{Bou99b}, which plays a central role
in the present work, generalizes a proposal by Fischler and
Susskind~\cite{FisSus98} and is thought to have its origin in the
holographic principle, first formulated by 't~Hooft~\cite{Tho93} and
Susskind~\cite{Sus95}.  The application of the covariant entropy bound
to the past light-cone of an observer was proposed by Banks in
Ref.~\cite{Ban00a}.  The D-bound is related to Bekenstein's bound on
the entropy of finite systems in flat space~\cite{Bek81}.  Its
derivation adapts the original arguments of Geroch and Bekenstein, and
extends to cosmologically large systems Schiffer's use of the
cosmological horizon to obtain Bekenstein's bound~\cite{Sch92}.
Bekenstein's generalized second law of
thermodynamics~\cite{Bek72,Bek73,Bek74} underlies most of the work in
this paper.  The semi-classical description of asymptotically
de~Sitter space-times was laid out by the work of Gibbons and
Hawking~\cite{GibHaw77a}; see also Ref.~\cite{FigHoe75}.  Other recent
work exploring connections between the holographic principle and the
cosmological constant includes Refs.~\cite{CohKap98,HorMin00,Tho00}.

\section{Causal diamonds}
\label{sec-cd}

We first address the question of which entropy (or information) is
actually accessible to a given observer.  We will argue that certain
space-time regions can be eliminated from consideration, and that the
$N$-bound need only hold for the remaining region, the `causal
diamond' associated with an observer.  It will also be shown that
these restrictions are necessary, in the sense that the inclusion of
unobservable entropy would easily allow the violation of the bound.

Implicit in this approach is the principle, long advocated by
Susskind, that a fundamental theory need only answer questions that
are operationally meaningful.  For example, it need not (and, from an
aesthetic standpoint, should not) simultanously describe the
experiments made by two separate observers who, for reasons of causal
structure, will never be able to compare results.  Of course, it must
be able to describe each experiment separately.  This principle has
previously been used to resolve certain apparent paradoxes in the
evaporation of black holes~\cite{SusTho93,Sus93,SusTho93b}.

We will consider an experiment that begins at point $p$ and ends at a
later point $q$ on the observer's world line.  It will be seen that
causality limits the space-time region whose entropy can play a role
in the experiment.  It may be sufficient to consider only `the longest
experiment possible', i.e., the limit in which $p$ is taken to be in
the far past, and $q$ in the far future, on the world line.  However,
it will be simpler and more instructive to carry out the discussion
for arbitrary $p$ and $q$.  As experiments often have finite duration,
this is the most general case; and all results will continue to hold
in the limit of early $p$ and late $q$.

\subsection{The past light-cone}

There are two independent restrictions.  The first is:
\[ \mbox{(R1)~~~\em
Consider only the observer's causal past, $J^-(q)$.  Ignore everything
else.}
\]
This is a sensible restriction.  At the point $q$, the endpoint of the
experiment, the observer can only have received signals from the past
of $q$.  The rest of space-time has not yet been seen.  For the
purposes of the experiment in question, its entropy is operationally
meaningless and can be ignored.

For the later application of entropy bounds, note that the observer's
past is bounded by the past light-cone from the point $q$, and that
all matter within the observer's past must pass through this cone.%
\footnote{This is intuitively obvious but can be made precise as
follows.  The causal past of $q$, $J^-(q)$, is defined as the set of
points that can be reached from $q$ via a smooth curve that is
everywhere past-directed timelike or null.  Define the past
light-cone, $L^-(q)$, as the hypersurface generated by the
past-directed null geodesics that start at $q$ and are terminated if
and only if they run into a point conjugate to $q$ (a `caustic').
Assuming global hyperbolicity one can show~\cite{Wald} that the
boundary of the past of $q$, $\dot{J}^-(q)$, is a portion of $L^-(q)$.
For any point $r \in J^-(q)$, we claim that all future inextendible
causal curves through $r$ must intersect $L^-(q)$.  In fact, the
stronger statement holds that they must intersect $\dot{J}^-(q)$; in
the notation of Wald~\cite{Wald}, $D^-[\dot{J}^-(q)] = J^-(q)$.  This
follows from the compactness of $J^+(r) \cap J^-(q)$ (Theorem 8.3.10)
and Lemma 8.2.1 in Wald~\cite{Wald}.}
Thus, if one wishes to bound the observable entropy, it will be
sufficient to bound the entropy on the past light-cone of the
endpoint, $q$.

\EPSFIGURE{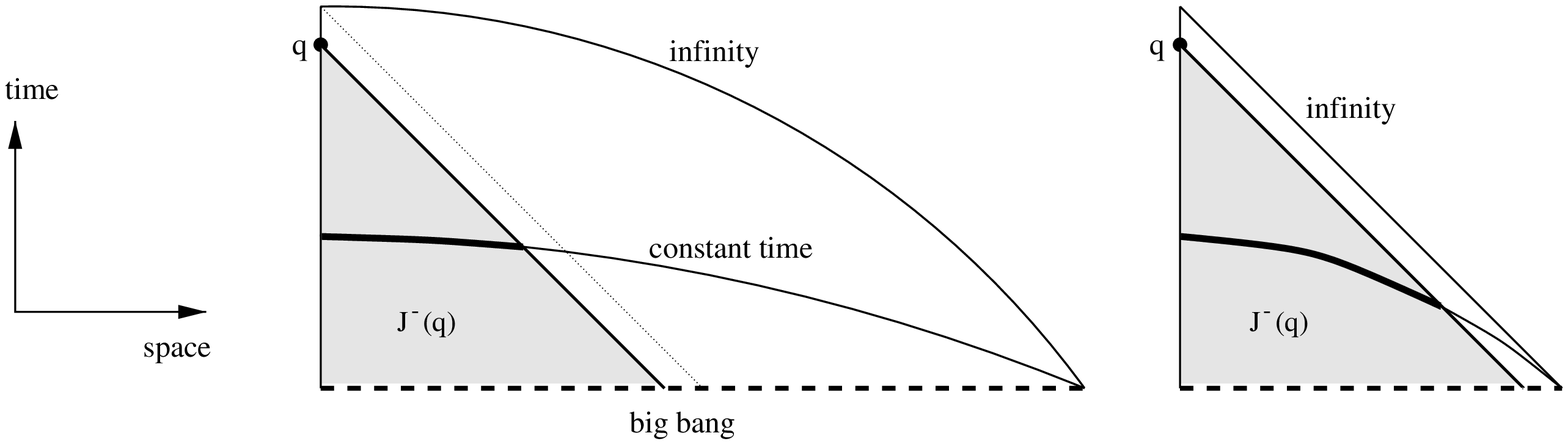,width=14cm}%
{Flat FRW universe with $\Lambda>0$ (left).  The entropy on any
constant-time slice is infinite, but only a finite portion (heavy
line) can be seen by the observer at $q$.  Because of the future
de~Sitter horizon (dotted line), this portion will not diverge.
Right: flat FRW universe with $\Lambda=0$.  The entropy within the
observer's past light-cone diverges at late times.%
\label{fig-R1}}
The restriction R1 is necessary for the $N$-bound.  Consider a
$\Lambda > 0$ flat FRW universe starting with a big bang---possibly a
good approximation to the universe we inhabit.  The entropy density on
any homogeneous spacelike slice is constant; thus, the total entropy
on the slice is formally infinite, in apparent violation of the
$N$-bound.  The restriction R1 resolves this problem.  Because the
cosmological constant dominates at late times, any observer has a
future event horizon (Fig.~\ref{fig-R1}).  The entropy in its
interior is finite.  Because the event horizon contains the observer's
past for any endpoint $q$, the observed entropy is also finite.  (We
do not show quantitatively that it satisfies the $N$-bound as this
will follow from the general arguments given in
Sec.~\ref{sec-nbound}.)

In this example, space-time is asymptotically de~Sitter in the future,
with entropy $N$.  Thus, R1 is not only sensible and necessary for the
$N$-bound, but indeed necessary for the validity of the generalized
second law of thermodynamics.

It is instructive to contrast the above example with the case of a
$\Lambda=0$ flat FRW universe (Fig.~\ref{fig-R1}).  The latter has a
different infinity structure.  Arbitrarily large portions of any flat
hypersurface lie within the past light-cone at sufficiently late
times.  Even with restriction R1, the observed entropy is unbounded.
Of course, this is not a problem, because $N = \infty$ in this case.

\subsection{The future light-cone}
\label{sec-futurecone}

The second restriction is:
\[ \mbox{(R2)~~~\em
Consider only the observer's causal future, $J^+(p)$.  Ignore
everything else.}
\]
Note that the observer's future is bounded by the future light-cone of
the point $p$, and that all matter within the observer's future must
have entered through this cone.%
\footnote{This follows by exchanging `past' and
`future', $-$ and $+$, and $q$ and $p$, in the previous footnote.}

This restriction may seem less obvious than the previous one.  But it
is just as sensible.  It is not enough for entropy, or information, to
lie in the observer's past.  To be observed, it actually has to get to
the observer, or at least to a region that can be probed by the
observer.  But an experiment that commences at $p$ can only probe what
is in the causal future of $p$.

\EPSFIGURE{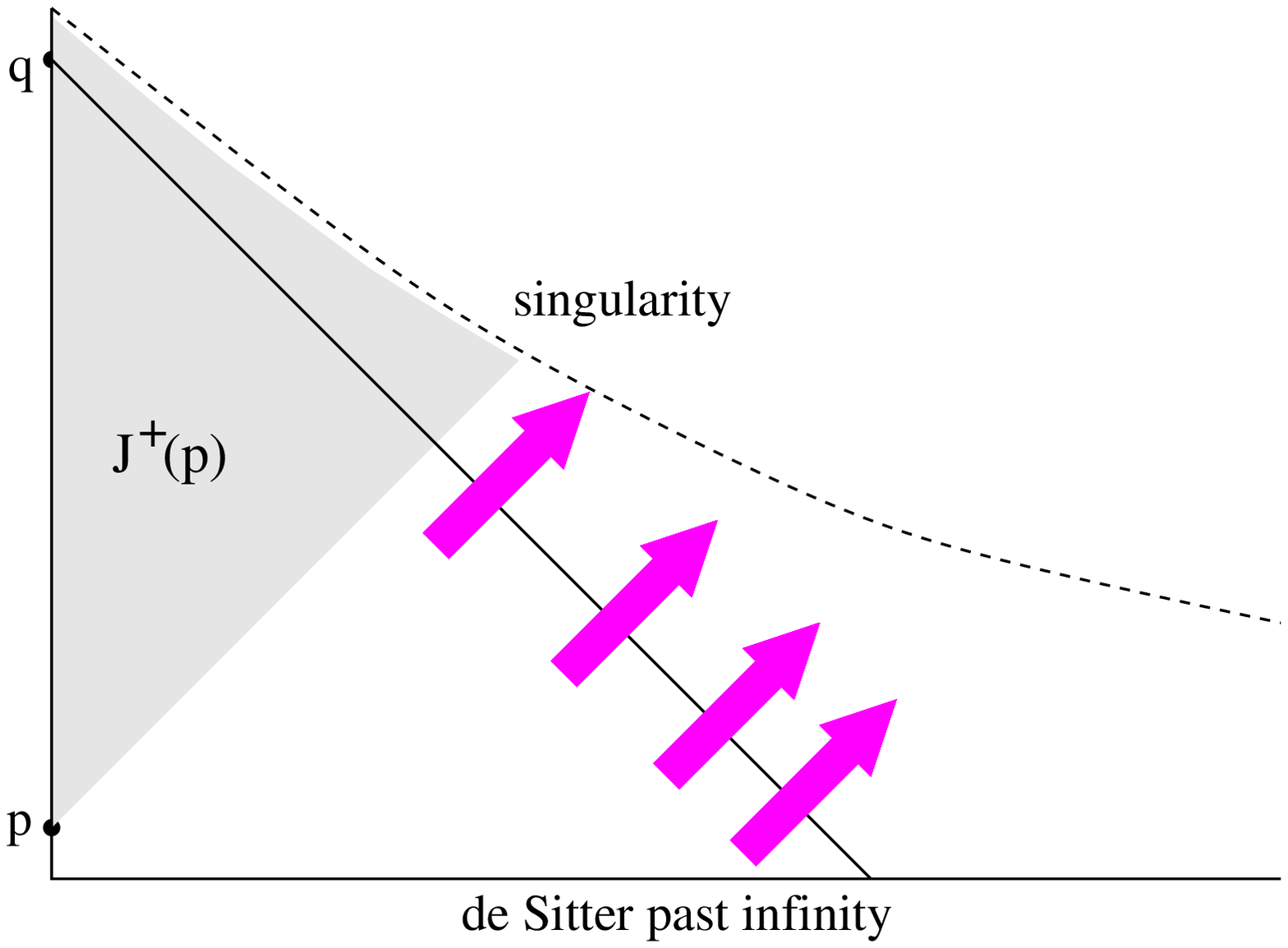,width=8cm}%
{A collapsing universe approaching de~Sitter in the past.  The past
light-cone of $q$ may contain infinite entropy (arrows), but only a
finite amount will enter the future of $p$ (shaded region).%
\label{fig-R2}}
Put differently, all information that reaches the observer, or at
least is accessible to the observer, must have \linebreak passed
through the future light cone of $p$.  For the purpose of describing
the experiment in question, one can ignore the space-time region
outside the cone; instead, one may think of the initial conditions as
residing on the cone.  Entropy that fails to enter through the cone is
operationally meaningless: \linebreak though it may well be present in
the observer's causal past, an experiment that starts at $p$ will not
know about it, because it cannot probe the region where such entropy
resides.

How is this consistent with cosmological observations of distant
galaxies?  By measuring the cosmic microwave background radiation, are
we not collecting information about the early universe?  These regions
are indeed outside the future of our entire world-line, let alone the
future of the point when the experiment began.  However, all the
information we gathered was in photons that interacted with some local
apparatus.  They had to enter through the future cone to get here.  So
the entropy we actually observe is quite local.  It is certainly
insightful to interpret this information in terms of models that
involve inaccessible regions.  For example, one might say that the
early universe contained certain density perturbations.  But the
information used to obtain this conclusion is here, now.  Thus, it is
subject to entropy bounds associated with a much smaller region than
the one it is interpreted to be an imprint of.

Without the additional restriction R2, the $N$-bound would fail.
Fig.~\ref{fig-R2} shows a $\Lambda > 0$ space-time in which the
observer's causal past contains an arbitrarily large entropy.
Consider a universe that approaches empty de~Sitter space
asymptotically in the past.  The geometry will resemble the lower half
of the de~Sitter hyperboloid at early times (see Appendix).  It
contains exponentially large three-spheres, on which one can place
dilute matter with arbitrary entropy.  If the total entropy exceeds
$N$, the universe will necessarily be dominated by this matter at a
later time.  It will collapse in a big crunch, and there will be no
future de~Sitter region.  One can arrange for the energy and entropy
density to be constant on the observer's past light-cone (by giving it
an increasing profile on the early $S^3$, in the radial direction away
from the observer's world line).  The past light-cone keeps going
forever, and so the total entropy on it will be infinite.---Note that
the area of surfaces on the past light-cone diverges, so this example
does not contradict the covariant entropy bound discussed in
Sec.~\ref{sec-ceb}.

\subsection{The causal diamond}

Recall that $p$ and $q$ are two points on an observer's world line,
with $q$ later than $p$.  One can think of $p$ as the beginning and
$q$ as the end of some experiment.  The restrictions R1 and R2 define
the space-time region that can come into play in such an experiment.
According to R2, one can ignore what is outside the causal future of
$p$, and R1 states that regions outside the causal past of $q$ are
operationally meaningless as well.  Combining both conditions, one can
restrict to the points which are both in the future of $p$ and in the
past of $q$.  This set,
\begin{equation}
C(p,q) = J^+(p) \cap J^-(q),
\end{equation}
will be called the {\em causal diamond\/} associated with an
experiment beginning at $p$ and ending at $q$.  Thus, one obtains the
condition
\begin{center}
\mbox{(R1+R2)~~~\em
Consider only the entropy in causal diamonds, i.e.,}
\mbox{~~~~~\em in regions of the
form $C(p,q)$.}
\end{center}
(The notion of an observer's world line was a crutch that can be
dropped now.  If $q$ is in the future of $p$, there will be world
lines connecting them; if not, then $C(p,q)$ will be empty or
degenerate.)

Of a fundamental theory, one may demand that it describe any
experiment, but no more than that.  Hence, it should describe the
physics in any causal diamond, that is, in any region of the form
$C(p,q)$ for some pair of points $(p,q)$, but {\em only one causal
diamond at a time}.  One should not demand that the theory
simultaneously describe two separate causal diamonds, unless they are
both contained in a single larger causal diamond.

For example, the theory should be able to describe an experiment
inside a black hole, as well as an experiment outside a black hole.
But it should not describe correlation functions between a point
inside and a point outside a black hole if those points do not lie in
any causal diamond.  This example is just a reformulation of some of
the arguments that established the concept of `black hole
complementarity'~\cite{SusTho93,Sus93,SusTho93b}.  (In this case only
the restriction R1 really matters, since R2 can easily be satisfied.)
An analogous argument can be made for pairs of points near a big bang
singularity.  If they are sufficiently far, they cannot lie in a
single causal diamond.  Then no experiment can be set up that will
involve both points.  (In this case, R2 is the crucial restriction.)

In space-times that are asymptotically de~Sitter in the past and
future, any causal diamond lies within both the past and future event
horizon.  (Both R1 and R2 are used here.)  The exponentially large
regions beyond those horizons are operationally meaningless.  This
result has long been advocated by Susskind.

\EPSFIGURE{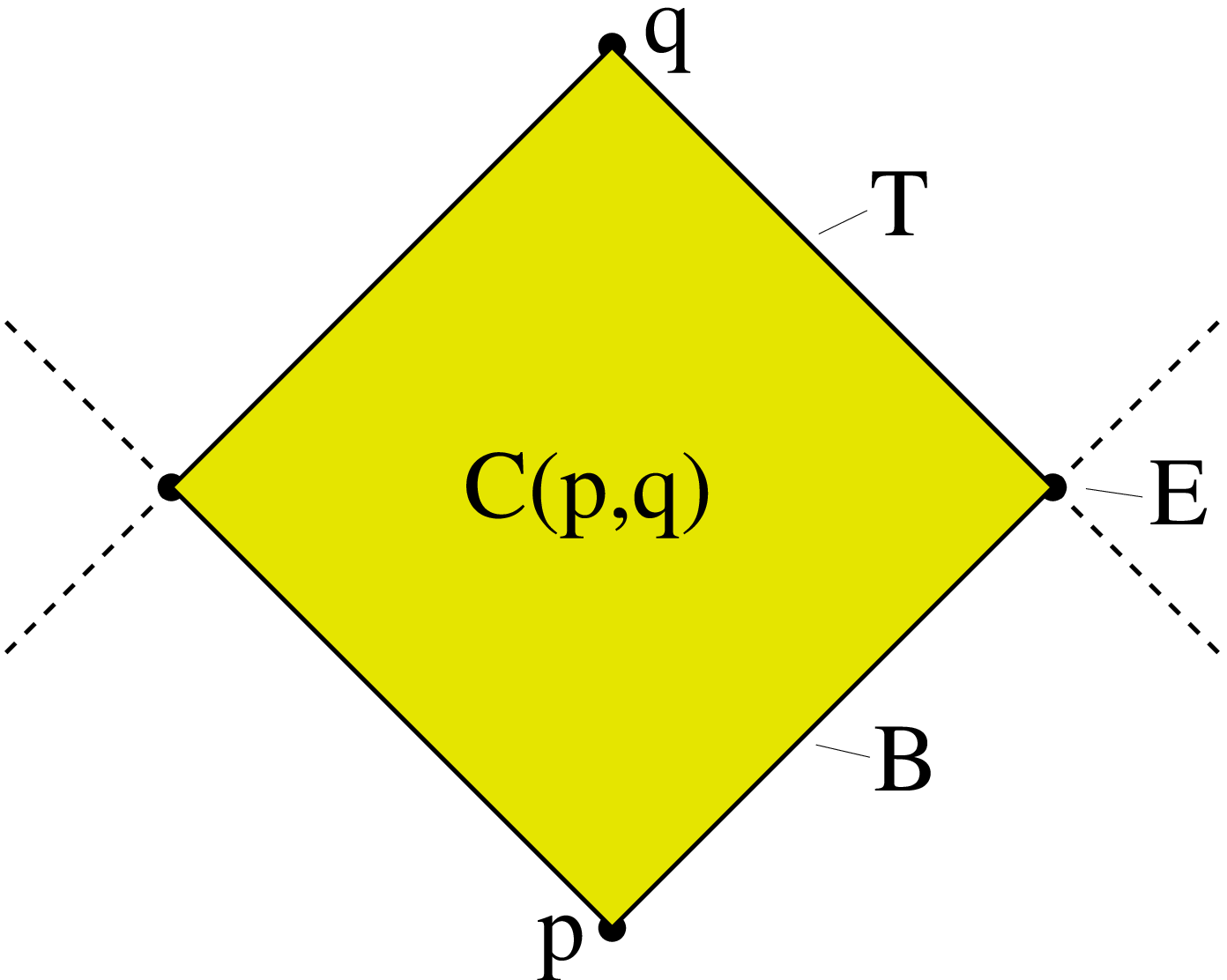,width=6cm}%
{A causal diamond, with top cone $T$, bottom cone $B$, and edge $E$.
\label{fig-cd}}
A causal diamond is bounded by a {\em top cone\/} (a portion of the
past light-cone of $q$), and a {\em bottom cone\/} (a portion of the
future light-cone of $p$); see Fig.~\ref{fig-cd}.  The cones usually,
though not necessarily, intersect at a two-dimensional spatial
surface, the {\em edge} of the causal diamond.  In any case, the
entropy in the causal diamond must pass through the top cone (and all
matter must have entered through the bottom cone).%
\footnote{This follows from the previous two footnotes, with $\dot
C(p,q) = [J^+(p) \cap \dot J^-(q)] \cup [\dot J^+(p) \cap J^-(q)]$.
The first term in square brackets is the top cone, $T(p,q)$; the
second is the bottom cone, $B(p,q)$; their intersection is the edge,
$E(p,q)$.  Clearly, $T(p,q) \subset \dot J^-(q) \subset L^-(q)$, and
similarly, $B(p,q) \subset L^+(p)$.}
It will be seen below that the nature of the boundaries allows for a
straightforward application of the covariant entropy bound.  For this
reason, the entropy within a causal diamond is under good theoretical
control.

\newpage

\section{The covariant entropy bound}
\label{sec-ceb}

The covariant entropy bound~\cite{Bou99b} bounds the entropy on
certain null hypersurfaces or `light-sheets'.  It was developed in
order to formulate the holographic principle~\cite{Tho93,Sus95} for
general space-times~\cite{Bou99c}, and can be viewed as a
generalization of the approach of Fischler and
Susskind~\cite{FisSus98}.  The use of null hypersurfaces to relate
entropy and area was originally suggested by Susskind~\cite{Sus95}.
Several concepts crucial to a general formulation were first
recognized by Corley and Jacobson~\cite{CorJac96}.

The bound is conjectured to hold for any spacial surface in any
space-time with reasonable energy conditions.  It will be useful here
because it applies even to regions, such as recollapsing universes or
black hole interiors, where the second law is of no help.  The
conjecture has passed a number of non-trivial tests~\cite{Bou99b}.  It
has been proven in space-time regions where a fluid approximation to
entropy can be made with plausible relations between entropy and
energy density~\cite{FlaMar99}.

\EPSFIGURE{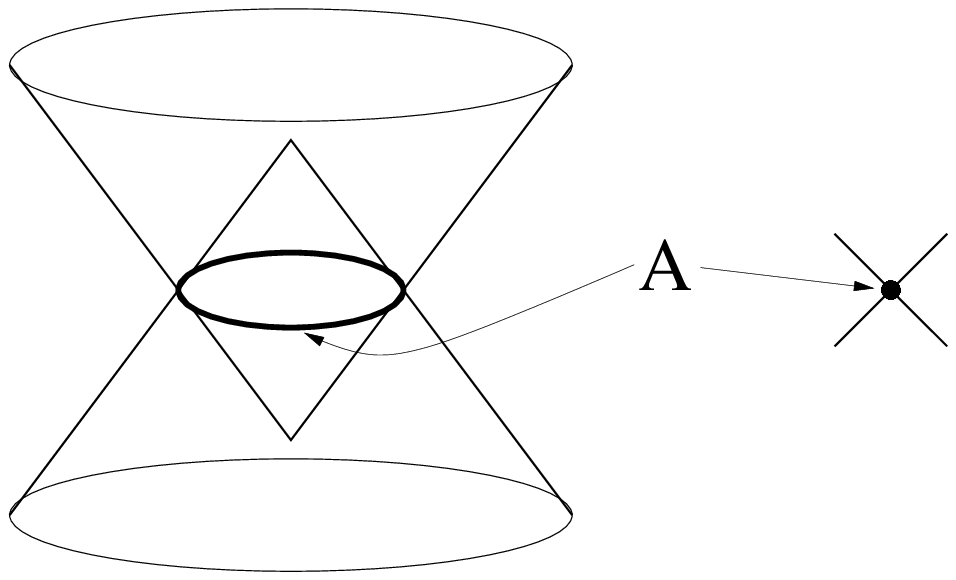,width=8cm}%
{The four light-like hypersurfaces orthogonal to a spatial surface (in
this example, two cones going in and two `skirts' going out).  In a
Penrose diagram the four null directions are indicated by an `X'
(right).
\label{fig-ceb1}}
Consider some 2-dimensional spatial surface of area $A$.  (We will
mostly be interested in closed surfaces, but this is not a necessary
restriction.)  Any surface has four orthogonal light-like directions.
Namely, the surface has two sides, and on each side there is a family
of orthogonal light-rays arriving from the past (past-directed
light-rays), and a family of future-directed light-rays.  In
Fig.~\ref{fig-ceb1} this is illustrated for the example of a spherical
surface.  In a Penrose diagram, where light travels at 45 degrees, the
four orthogonal light-like directions are indicated by the legs of an
`X' centered on the point that represents the sphere.

The orthogonal light-rays generate four 2+1 dimensional null
hypersurfaces.  On some of them, the {\em light-sheets\/} of the
surface $A$, the cross-sectional area spanned by the light-rays will
be decreasing or constant in the direction away from the original
surface.  (In the example in Fig.~\ref{fig-ceb1}, the two cones.)  The
entropy on any light-sheet is less than $A/4$:
\begin{equation}
S(\mbox{light-sheet of $A$}) \leq \frac{1}{4} A.
\end{equation}

Any surface has at least two light-sheets, since two of the four
families of light-rays are just continuations of the opposite two.
E.g., if the area is increasing in the future direction to one side,
it will necessarily decrease in the past direction to the other side.
If it is constant in some direction, both opposing legs will be
allowed.

\EPSFIGURE{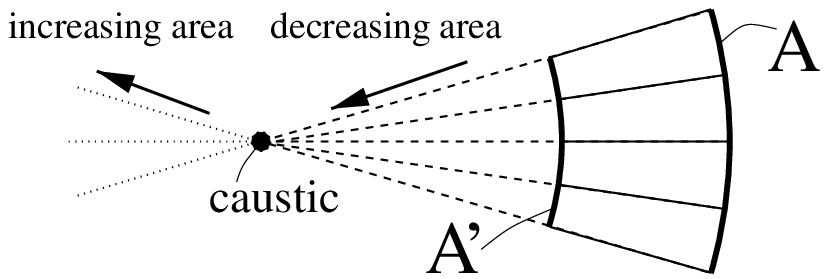,width=8cm}%
{A light-sheet is a null hypersurface with shrinking cross-sectional
area.  It terminates at caustics (if not earlier).
\label{fig-ceb2}}
The requirement of decreasing cross-sectional area is a local
condition and it must hold everywhere on the light-sheet.  This means
that the light-sheet must be terminated at or before one reaches a
caustic, i.e., before neighbouring light-rays intersect
(Fig.~\ref{fig-ceb2}).  In Fig. \ref{fig-ceb1}, the tips of the cones
are caustics, and the light-sheets end there.  The focussing theorem
guarantees that contracting light-rays will not become expanding
without going through a caustic.%
\footnote{The null convergence condition~\cite{HawEll} is assumed to
hold: $T_{ab} k^a k^b \geq 0$ for all null vectors $k^a$. --- It has
been suggested~\cite{TavEll99} that a light-sheet be terminated also
at points where non-neighbouring light-rays intersect.  As this can
only make the light-sheet smaller, it gives a weaker bound, but the
smaller light-sheet may be easier to compute practically.  The
light-sheets in Sec.~\ref{sec-nbound} below are of this simple type,
because they are a portion of the boundary of the causal past of a
point.} %
If one chooses to terminate the light-sheet before each light-ray
reaches a caustic, the end-points will span a non-zero area $A'$.
Then the covariant bound can be strengthened~\cite{FlaMar99}:
\begin{equation}
S \leq \frac{1}{4} \left( A - A' \right).
\end{equation}

The light-sheet directions associated with a surface can be indicated,
in a causal diagram, by the corresponding legs of the `X'
(Fig.~\ref{fig-ceb3}).  The two allowed directions form a wedge.  One
may classify closed surfaces as follows.  For {\em normal\/} surfaces,
both legs of the wedge point to one side, which is called the {\em
inside\/} by definition.  If both light-sheets are future-directed,
the surface is {\em trapped\/}; if the area is contracting in both
past directions, it is called {\em anti-trapped\/}.  Marginal cases
arise for surfaces on the interface between a normal and a trapped or
anti-trapped region.  Then the expansion vanishes along at least one
opposing pair of legs, and three or four legs must be drawn.  The
covariant entropy bound is particularly powerful when applied to such
surfaces, and we will focus on them in Sec.~\ref{sec-nbound}.
\EPSFIGURE{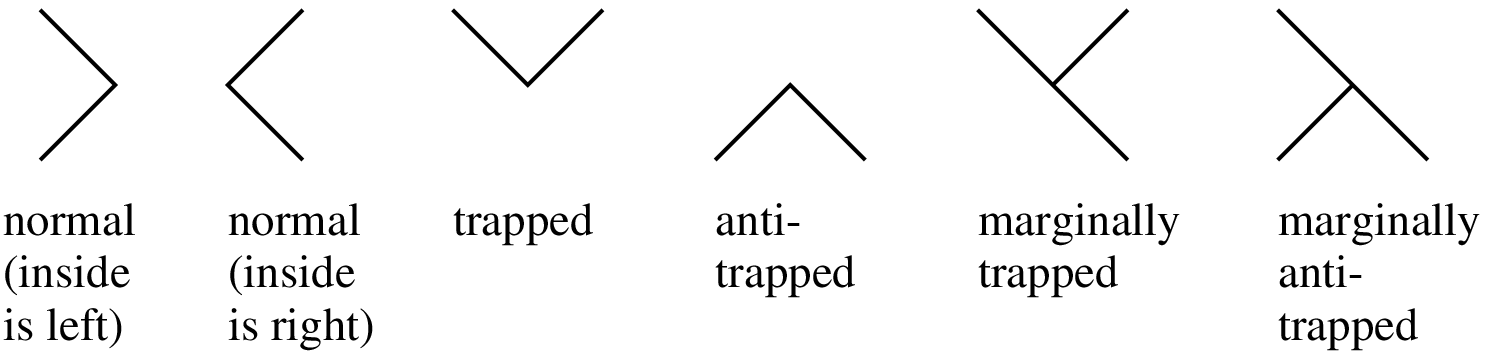,width=12cm}%
{Wedge symbols for different types of surfaces.  A leg is drawn for
each direction in which light-rays are non-expanding.
\label{fig-ceb3}}

\section{The D-bound on matter entropy in de~Sitter space}
\label{sec-dbound}

By studying the second law of thermodynamics in asymptotically flat
space, Bekenstein found that the total entropy is given by the sum of
ordinary matter entropy, $S_{\rm m}$, plus the semiclassical
Bekenstein-Hawking entropy, $S_{\rm h} = \frac{1}{4} A_{\rm h}$,
associated with the horizons of black
holes~\cite{Bek72,Bek73,Bek74,Haw74}.  Similarly, in asymptotically
de~Sitter space, the cosmological horizon contributes with
\begin{equation}
S_{\rm c} = \frac{1}{4} A_{\rm c}
\end{equation}
to the total entropy~\cite{GibHaw77a}.

Empty de~Sitter space has a cosmological horizon of area
\begin{equation}
A_0 = \frac{12\pi}{\Lambda} = 4N
\end{equation}
(see Appendix).  Therefore, empty de~Sitter space has horizon entropy
$S_{\rm c} = N$.  One might think that even a tiny amount of matter
entropy would already increase the total entropy, $S_{\rm c} + S_{\rm
m}$, above $N$.  However, the cosmological horizon surrounding a
matter system in asymptotically de~Sitter space is smaller than $A_0$:
the more matter, the smaller the cosmological horizon.  Thus it is
possible that the total entropy remains bounded by $N$.  (It is
instructive to verify this explicitly for the simple case of
Schwarzschild-de~Sitter black holes; see also~\cite{MaeKoi98}.)

It will now be shown that the $N$-bound is in fact implied by the
second law {\em if} space-time contains an asymptotically de~Sitter
region in the future.  This will allow us, by subtracting the horizon
entropy from $N$, to derive a bound on the matter entropy in de~Sitter
space.  Despite the restrictive assumption of an asymptotic de~Sitter
region, this bound will be useful to our purpose; we will argue later
that it may also be applied to certain portions of more general
space-times.  Thus it will join the covariant bound, and the concept
of causal diamonds, as a third ingredient in the argument constructed
in Sec.~\ref{sec-nbound} to show that the $N$-bound is valid for all
$\Lambda>0$ space-times.

Consider the following process.  The initial configuration is a matter
system in asymptotically de~Sitter space.  The matter system may
contain black holes, whose entropy is taken to be included in the
matter entropy, $S_{\rm m}$.  The system is surrounded by a
cosmological horizon of area $A_{\rm c}$.  The final state is empty
de~Sitter space.  The transition is achieved by taking the observer to
move into the asymptotic region.  (To the observer, the matter system
appears to fall into the cosmological horizon.)  In this process, the
matter entropy $S_{\rm m}$ is lost, while the entropy of the
cosmological horizon increases by an amount
\begin{equation}
\Delta S_{\rm c} = \frac{1}{4} \left( A_0-A_{\rm c} \right).
\end{equation}

The generalized second law of
thermodynamics~\cite{Bek72,Bek73,Bek74} implies that the total entropy
must not decrease:
\begin{equation}
\Delta S_{\rm c} \geq S_{\rm m}.
\end{equation}
With $A_0 = 4N$, one obtains a bound on the matter entropy:
\begin{equation}
S_{\rm m} \leq N - \frac{1}{4} A_{\rm c}.
\label{eq-Dbound}
\end{equation}
To distinguish this bound from the covariant entropy bound and the
$N$-bound, it will be called the {\em D-bound\/} (`D' as in Difference
between $N$ and the horizon entropy).

The D-bound is less general than the covariant bound of
Sec.~\ref{sec-ceb}, because it only applies to matter systems within a
de~Sitter horizon.  For a dilute system, one has $A_{\rm c} \approx
A_0$, and therefore, $ N - \frac{1}{4} A_{\rm c} \ll \frac{1}{4}
A_{\rm c}$.  So the D-bound can be tighter than the covariant bound
applied to a surface enclosing the system.  In the next section it
will be seen that causal diamonds can contain portions to which the
D-bound applies.

\section{The $N$-bound}
\label{sec-nbound}

In Sec.~\ref{sec-intro} the $N$-bound was presented as a conjecture:
The observable entropy in any $\Lambda>0$ universe cannot exceed $N =
3\pi/\Lambda$.  In Sec.~\ref{sec-cd} it was shown that only the
entropy within space-time regions of a particular form, causal
diamonds, is observable.  Hence, to prove the $N$-bound, it suffices
to show that the entropy of an arbitrary causal diamond does not
exceed $N$.  By applying the covariant entropy bound
(Sec.~\ref{sec-ceb}) and the D-bound (Sec.~\ref{sec-dbound}), we will
now give a proof for spherically symmetric causal diamonds.  Spherical
symmetry allows us to keep the discussion fairly non-technical and
focus on the key idea, the interplay between the D-bound and the
covariant bound.  We expect that the assumption of spherical symmetry
can be eliminated in a more refined treatment; this will be discussed
briefly at the end of the section.

Consider an experiment beginning at a point $p$ and ending at $q$, in
a universe with $\Lambda > 0$.  We must show that the matter entropy,
$S_{\rm m}$, within the causal diamond, $C(p,q)$, plus the
Bekenstein-Hawking entropy of any black hole or cosmological horizons
identified by the experiment, will not exceed $N$.  To limit $S_{\rm
m}$, it suffices to consider the matter entropy passing through the
top cone bounding the diamond, by the second law.  It will be seen
that the horizon entropy is bounded by the area of the diamond's edge.

Neither the top nor the bottom cone contain any caustics, except at
endpoints, because each is a portion of the boundary of the future or
past of a point.  Then, by the focussing theorem, each cone has
exactly one maximal cross-sectional area.  The maximum may be local,
or it may lie on the intersection of the two cones, the edge, where
they terminate.  Depending on the location of the maxima, we
distinguish three cases.

\paragraph{Case 1: No local maximum on either cone.}
Then the maximum area of each cone lies on the edge.  Thus, the edge
will be a normal surface, with the observer on the inside
(Fig.~\ref{fig-case1}).  This case applies, for example, to regions
within the horizon in an asymptotically de~Sitter universe, and to
sufficiently small causal diamonds in arbitrary spacetimes.

\EPSFIGURE{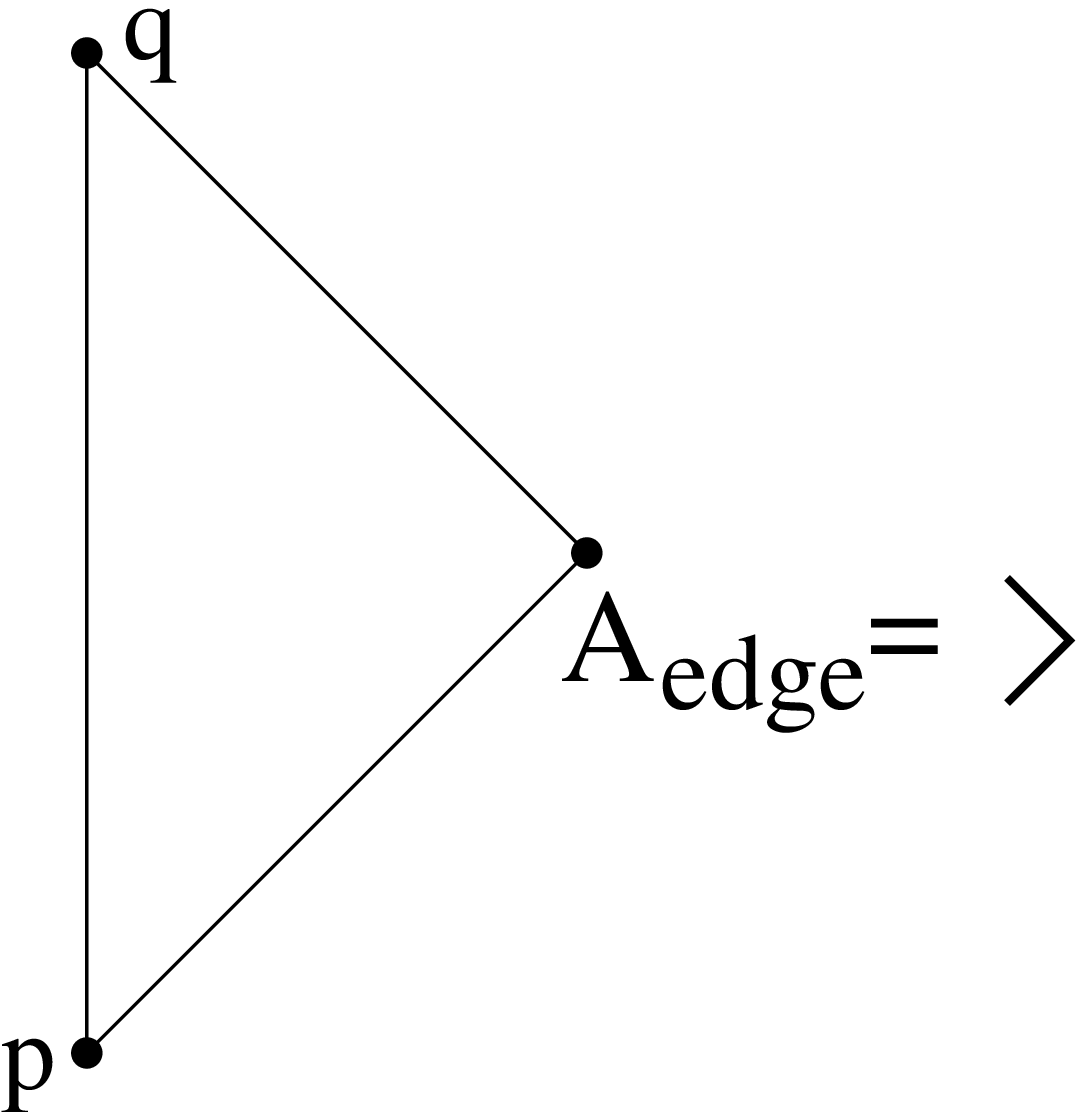,width=4.5cm}%
{Causal diamond in Case 1 (to be rotated about the $p$-$q$ axis).  The
edge is normal.  The D-bound applies to the top cone.
\label{fig-case1}}
Consider a space-like hypersurface containing the edge.  One can
consistently assume, for the sake of argument, that the exterior of
the edge is a vacuum solution.  With the assumption of spherical
symmetry, Birkhoff's theorem implies that the exterior will be a
portion of a Schwarzschild-de~Sitter (or a
Reissner-Nordstr\"om-de~Sitter) solution.  The space-time thus
constructed will be called the {\em auxiliary space-time}.  It is
asymptotically de~Sitter in the future and past; hence, it invites the
application of the D-bound.

The causal diamond lies within the cosmological horizon of the
auxiliary space-time, because the edge is normal and the cosmological
horizon is the outermost normal surface.  With spherical symmetry it
follows that $\hat{A}_{\rm c} \geq A_{\rm edge}$.  The hat indicates
that the cosmological horizon is a surface in the auxiliary
space-time.  Because the auxiliary space-time is asymptotically
de~Sitter, the D-bound can be applied to the interior of the
cosmological horizon, yielding
\begin{equation}
S_{\rm m} \leq N - \frac{1}{4} \hat A_{\rm c}
          \leq N - \frac{1}{4} A_{\rm edge}.
\end{equation}

Recall from Sec.~\ref{sec-dbound} that the entropy of black holes is
already subsumed in $S_{\rm m}$,%
\footnote{This may play a role when spherical symmetry is abandoned.
In the presence of black holes, the edge can contain additional
disconnected components, namely spherical surfaces surrounding the
black hole horizons, within the cosmological horizon.} %
but not the entropy of the cosmological horizon (supposing that one
exists in the actual space-time under consideration).  However, the
observer cannot assign more cosmological horizon entropy than a
quarter of the area of the outermost surface that has been probed:
\begin{equation}
S_{\rm c} \leq \frac{1}{4} A_{\rm edge}.
\end{equation}
The two inequalities imply the $N$-bound,
\begin{equation}
S = S_{\rm m} + S_{\rm c} \leq N.
\end{equation}

\paragraph{Case 2: Local maximum on the top cone.}
Now assume that the top cone contains a locally maximal area $A_{\rm
max}$, an {\em apparent horizon\/}.  No assumption is made about the
bottom cone.  Cases of this type include large expanding or collapsing
cosmological regions.  They correspond to highly dynamical situations
without a quasi-static cosmological horizon, so $S_{\rm c}=0$.  Then
we need to show only that the matter and black hole entropy on the top
cone does not exceed $N$.

\EPSFIGURE{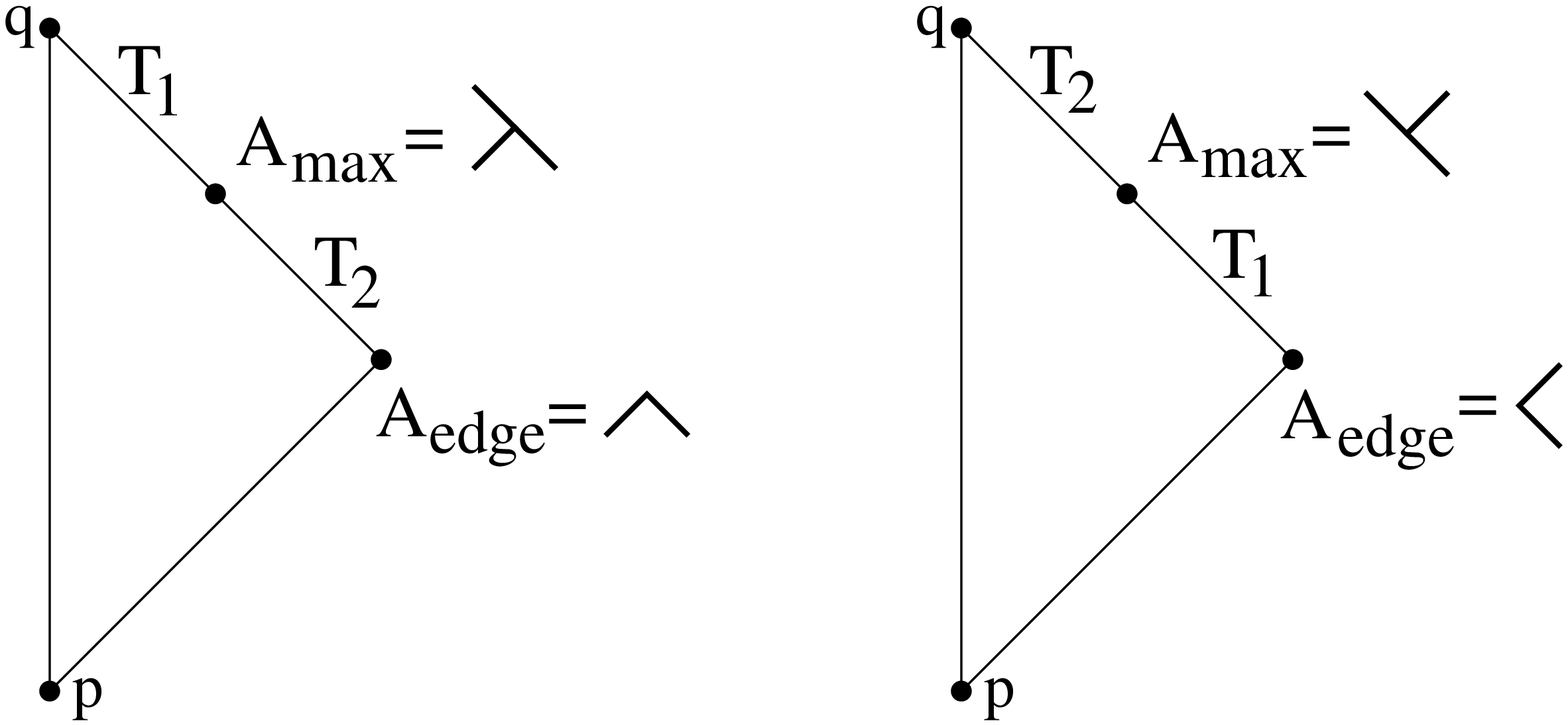,width=12cm}%
{Case 2.  One side of the maximal area must be normal.  We apply the
D-bound to this side, $T_1$, and the covariant bound to the other
side, $T_2$.  Both possibilities are shown.
\label{fig-case2}}
The maximal area divides the top cone into two parts.  We will show
that the D-bound can be applied to one part and the covariant bound to
the other.  Recall the wedge formalism summarized in
Fig.~\ref{fig-ceb3}.  The wedge of the surface $A_{\rm max}$ is
constructed by drawing a leg for each light-like direction with
decreasing cross-sectional area.  Because $A_{\rm max}$ is the largest
surface on the top cone, the area obviously decreases in the two null
directions that generate the cone.  Of the other two null directions
orthogonal to $A_{\rm max}$, at least one must have decreasing area,
because they oppose each other.  Hence, the wedge associated with
$A_{\rm max}$ has at least three legs.  Necessarily, two of them will
be pointing to the same spatial side (Fig.~\ref{fig-case2}).
Therefore, $A_{\rm max}$ is a marginally normal surface.  The side
with two legs is, in the wedge sense, the inside of $A_{\rm max}$.
The corresponding portion of the top cone will be called $T_1$.  Note
that $T_1$ need not be the portion that includes the tip; it may be on
the `far side' of $A_{\rm max}$ (Fig.~\ref{fig-case2}, right).

Consider the inside portion, $T_1$, in isolation.  To this
hypersurface one can apply the D-bound, using an argument similar to
that of Case 1.  One can take $T_1$ to be embedded in an otherwise
vacuous auxiliary space-time.  Because $T_1$ is the interior of a
normal surface, in the auxiliary space-time it will be surrounded by a
cosmological horizon.  The area of the cosmological horizon will be no
less than $A_{\rm max}$.  Because the auxiliary space-time is
asymptotically de~Sitter, the D-bound applies to the interior of the
cosmological horizon.  Hence, the entropy on $T_1$ satisfies
\begin{equation}
S_1 \leq N - \frac{1}{4} \hat A_{\rm c}
    \leq N - \frac{1}{4} A_{\rm max}.
\end{equation}

The other part, $T_2$, of the top cone, is a light-sheet of the
surface $A_{\rm max}$.  The covariant entropy bound yields
\begin{equation}
S_2 \leq \frac{1}{4} A_{\rm max}.
\label{eq-S_2}
\end{equation}
It follows that the entropy on the top cone is bounded by $N$:
\begin{equation}
S = S_1 + S_2 \leq N.
\label{eq-Scase2}
\end{equation}

In this result, $S$ already includes black hole entropy.  A horizon is
probed by an experiment only if the edge of the causal diamond
contains a portion in the vicinity of the horizon.  The edge lies on
the far side of the top cone.  If this is $T_1$, the side to which the
D-bound applies, then the black hole entropy is already subsumed in
$S_1$, as discussed in Sec.~\ref{sec-dbound}.  If the far side is
$T_2$, let us split $S_2$ into black hole horizon entropy, $S_{\rm
h}$, and ordinary matter entropy, $S_{\rm m}$:
\begin{equation}
S_2 \equiv S_{\rm h} + S_{\rm m}.
\end{equation}
If $A_{\rm edge}>0$, the covariant bound on
$T_2$ can be strengthened~\cite{FlaMar99}:
\begin{equation}
S_{\rm m} \leq \frac{1}{4} \left( A_{\rm max} - A_{\rm edge} \right).
\end{equation}
The horizon cannot be larger than the area of the edge:
\begin{equation}
S_{\rm h} \leq \frac{1}{4} A_{\rm edge}.
\end{equation}
So Eq.~(\ref{eq-S_2}) holds, and $S$ in Eq.~(\ref{eq-Scase2}) is
indeed the total observable entropy.

\paragraph{Case 3: Local maximum on the bottom cone but
 not on the top cone.}  Finally, consider the case where the top cone
has no local maximum, but the bottom cone does (Fig.~\ref{fig-case3}).
Examples include large regions in collapsing universes, or black hole
interiors.  The edge of the cone is a trapped surface in this case.
This implies a dynamical situation without Bekenstein-Hawking entropy.
It will suffice to show that the matter entropy on the top cone does
not exceed $N$.

\EPSFIGURE{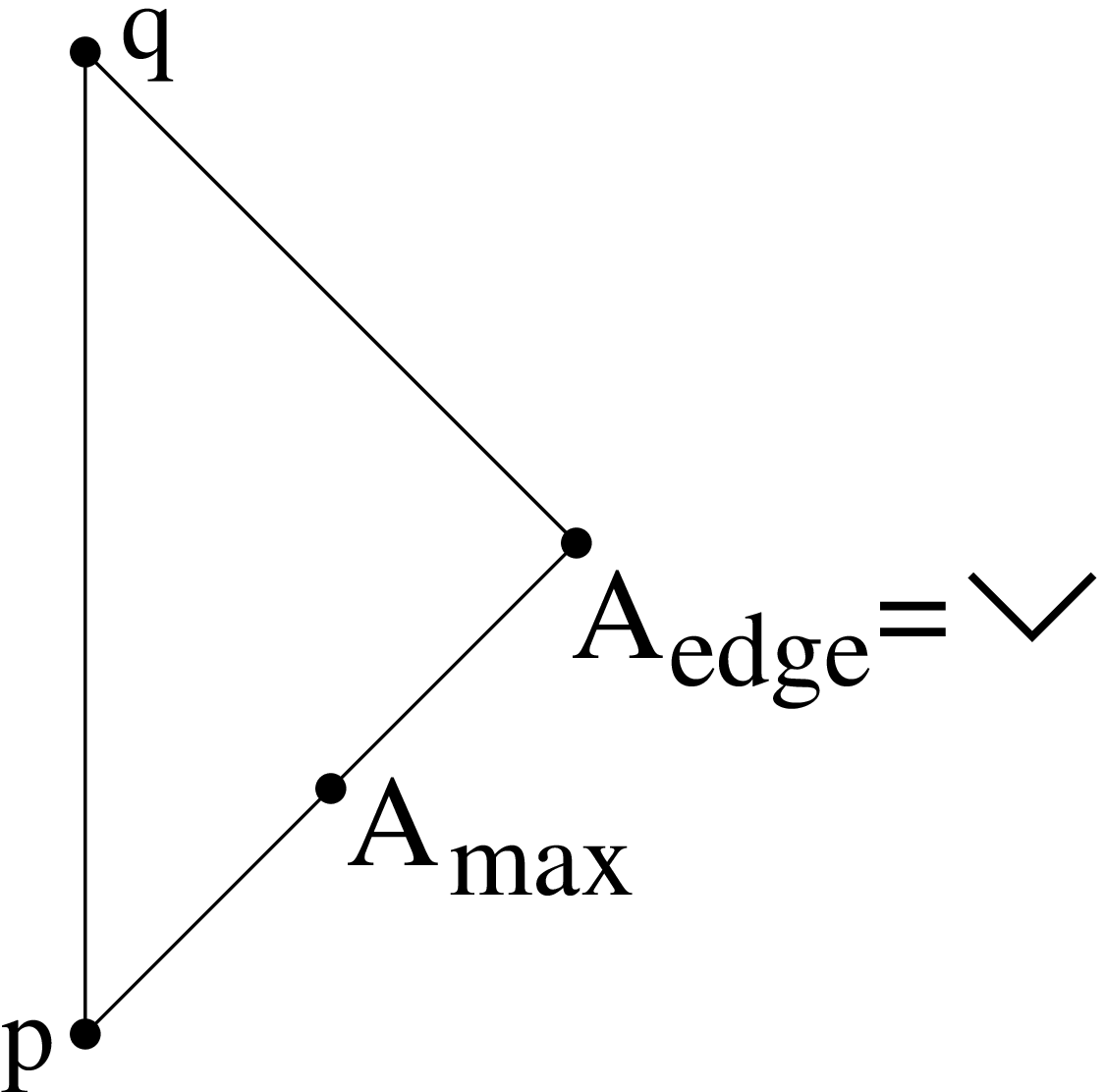,width=4.5cm}%
{Case 3.  The edge is trapped.  We apply only the covariant entropy
bound.
\label{fig-case3}}
In the absence of a local maximum, the largest surface on the top cone
is the edge.  The entire top cone is a light-sheet of the edge.  By
the covariant entropy bound,
\begin{equation}
S \leq \frac{1}{4} A_{\rm edge}.
\end{equation}
By the arguments used in Case 2, the maximal area on the bottom cone,
$A_{\rm max}$, is a normal surface.  Hence, it can be embedded in an
asymptotically de~Sitter auxiliary space time, where it is surrounded
by a cosmological horizon of area $\hat A_{\rm c}$.  By the second law,
$\hat A_c$ cannot exceed the horizon area of empty de~Sitter space,
$A_0$.  Moreover, by construction, $A_{\rm max}$ is larger than the
edge.  In summary, one finds
\begin{equation}
A_{\rm edge} \leq A_{\rm max} \leq \hat A_{\rm c} \leq A_0 = 4N.
\end{equation}
Therefore the $N$-bound is satisfied:
\begin{equation}
S \leq N
\end{equation}

In all three cases, we have used an auxiliary construction by which
(portions of) the causal diamond were embedded in an asymptotically
de~Sitter auxiliary space-time.  This method is rigorous only for
spherically symmetric situations.  The assumption was used in applying
Birkhoff's theorem to establish the auxiliary space-time, and in
taking the cosmological horizon as an upper bound on the area of
normal surfaces on the light-cone.  Spherical symmetry has also
simplified the case distinction, since it implies that the maximum is
either local or entirely on the edge; in general, the maximal area on
the top cone may have locally maximal components as well as portions
that lie on the edge.

Our assumption of spherical symmetry notwithstanding, we expect that
the above arguments represent the core of a general proof.  Causal
diamonds, light-sheets, and the entropy bounds are all defined without
reference to spherical symmetry.  The task of combining them to derive
the $N$-bound in the non-spherical case is left to future work.

\section{Outlook}
\label{sec-summary}

Non-perturbative definitions of quantum gravity have been given for
certain space-times that are asymptotically flat or
AdS~\cite{BanFis96,Mal97}.  No such description has been found for
space-times with a positive cosmological constant.  As no de~Sitter
solutions of M-theory are known, one does not even have a microscopic
framework.  Banks~\cite{Ban00} has opened a new perspective on this
problem by suggesting that an asymptotically de~Sitter universe is
described by a microscopic theory with finite-dimensional Hilbert
space.  Quantitatively, the $\Lambda$-$N$ correspondence relates the
cosmological constant of a stable vacuum, $\Lambda$, to a theory with
$N=3\pi/\Lambda$ degrees of freedom (i.e., with a Hilbert space of
dimension $e^N$).

If this is correct, M-theory (as it is currently understood) will
arise only in the limit of vanishing $\Lambda$ and infinite $N$.  The
cosmological constant problem becomes a problem of understanding the
particular dimension of Hilbert space chosen for the theory.

In Sec.~\ref{sec-banks}, considerations of consistency with
semi-classical gravity led us to propose the stronger conjecture that
the $\Lambda$-$N$ correspondence applies to all universes with
$\Lambda>0$, whether they are de~Sitter in the future or not.  This
conjecture makes the non-trivial, testable prediction that the
observable entropy in all such universes is bounded by $N$.  We then
argued that this statement, the `$N$-bound', is correct.  This
required the combination of the covariant entropy bound with two
intermediate results derived in Secs.~\ref{sec-cd}
and~\ref{sec-dbound}: the D-bound, and the restriction to causal
diamonds.  It is hard to see what, other than the $\Lambda$-$N$
correspondence, would offer a compelling explanation why such
disparate elements appear to join seamlessly to imply a simple and
general result.

The D-bound has a number of properties that merit further
investigation.  In particular, one can show that it is closely related
to Bekenstein's bound~\cite{Bek81}.  Bekenstein's bound, valid for
systems in flat space, can be written as $S_{\rm m} \leq \pi r_{\rm g}
R$, where $r_{\rm g} = 2m$ is the `gravitational radius' of the system
and $R$ is the circumscribing radius.  For dilute, spherically
symmetric systems in de~Sitter space, the D-bound takes precisely this
form as well, despite the significant deviation from flat space.  A
full discussion is given elsewhere~\cite{Bou00b}.

The restriction to causal diamonds arose in Sec.~\ref{sec-cd} from the
requirement to include only operationally meaningful parts of a
space-time in a microscopic description.  This principle is
independent of the present context of positive $\Lambda$, and one may
expect that causal diamonds will be of wider use.  Banks~\cite{Ban00a}
has sketched a framework for the combination of quantum mechanics and
cosmology, in which the variable size of the quantum Hilbert space is
related to the maximal area of the observer's past light-cone.  The
arguments of Sec.~\ref{sec-futurecone} suggest a possible modification
of this approach that may lead to a more time-symmetric treatment
based on the Hilbert space of causal diamonds.

In de~Sitter space, an observer will be immersed in quantum radiation
coming from the cosmological horizon.  At the semi-classical level,
this radiation is thermal~\cite{GibHaw77a}.  One would expect that the
radiation will occasionally contain large fluctuations that appear to
an observer as classical objects.  Taking a global view of the
de~Sitter space, one would say that quantum fluctuations originate
behind the future horizon, while classical objects enter through the
past horizon.  When one restricts to causal diamonds, however, both of
these outside regions are eliminated.  Then it is no longer clear how
an observer can distinguish between a genuine classical object and a
fluctuation in the quantum radiation.  (This view has previously been
advocated by Susskind.)  Indeed, all of standard cosmology may be a
rare fluctuation in a long-lived de~Sitter space~\cite{Ban00}.

Most of high energy physics is based on the S-matrix, with the
implicit assumptions that an observer of infinite size is located at
the infinity of an asymptotically flat space-time---the observer is
`outside looking in'.  This point of view will have to be transcended
in order to describe experiments in cosmology, where the observer is
always of finite size, and is `inside looking out'.  Indeed, a
Minkowski infinity typically does not exist in cosmology; but even if
it did, real observers would not live there.  On the other hand, the
approximation of an observer as a point in the space-time bulk is also
unsatisfactory, because an experiment involves the collection and
analysis of information.  According to the holographic principle, a
non-vanishing amount of information can be obtained only by an
observer of non-zero size.  The maximal information involved in an
experiment is related not to the size of the universe, but to the size
of the experiment.  One may be motivated by these considerations to
abandon the distinction between observer and experiment, and also to
claim that a general experiment {\em is} a causal diamond.  The bottom
cone is best thought of as arising from the limitation of preparing
the apparatus in a causal way; the future cone reflects the limitation
of analysing the data causally.

The correspondence between finite $N$ and positive $\Lambda$ would
impose a surprisingly strong restriction on the fundamental theory, if
indeed we live in a universe with positive (and true) vacuum energy.
We believe that its implications deserve to be further explored.

\acknowledgments

I have benefitted from discussions with many colleagues, especially
T.~Banks, J.~Bekenstein, O.~DeWolfe, M.~Fabinger, W.~Fischler,
V.~Hubeny, A.~Mikhailov, H.~Ooguri, S.~Shenker, L.~Susskind, R.~Wald,
and E.~Witten.  This research was supported in part by the National
Science Foundation under Grant No.\ PHY99-07949.

\appendix
\section{de Sitter space}

This Appendix summarizes a number of properties of de~Sitter
space that are used in the text.  An excellent discussion of the
classical geometry is found in Ref.~\cite{HawEll}.  The semi-classical
properties are laid out in Ref.~\cite{GibHaw77a}.

\EPSFIGURE{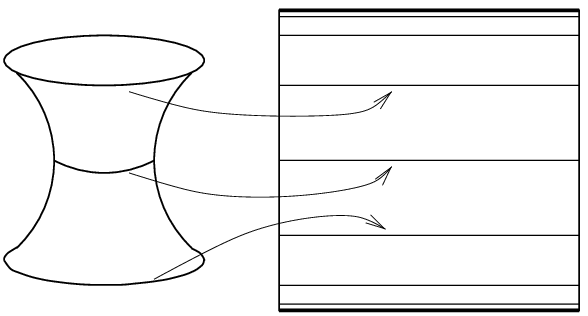,width=8cm}%
{de Sitter space as a hyperboloid.  Right: Penrose diagram.
Horizontal lines represent three-spheres.%
\label{fig-closed}}
de~Sitter space is the maximally symmetric solution of the vacuum
Einstein equations with a positive cosmological constant, $\Lambda$.
It is positively \linebreak curved with characteristic length
\begin{equation}
r_0  = \sqrt{\frac{3}{\Lambda}}
\end{equation}
Globally, de~Sitter space can be written as a closed FRW universe:
\begin{equation}
ds^2 = - dT^2 + r_0^2 \left( \cosh \frac{T}{r_0} \right)^2
 d\Omega_3^2
\end{equation}
The spacelike slices are three-spheres.  The space-time can be
visualized as a hyperboloid~\cite{HawEll} (Fig.~\ref{fig-closed}).
The smallest $S^3$ is at the throat of the hyperboloid, at $T=0$.  For
$T>0$, the three-spheres expand exponentially without bound.  The time
evolution is symmetric about $T=0$, so three-spheres in the past are
arbitrarily large and contracting.

The Penrose diagram of de~Sitter space is a square
(Fig.~\ref{fig-closed}).  The spatial three-spheres are horizontal
lines.  As usual, every point represents a two-sphere, except the
points on the left and right edge of the square, which represent the
poles of the three-sphere.  The top and bottom edge are the future and
past infinity, where all spheres become arbitrarily large.

\EPSFIGURE{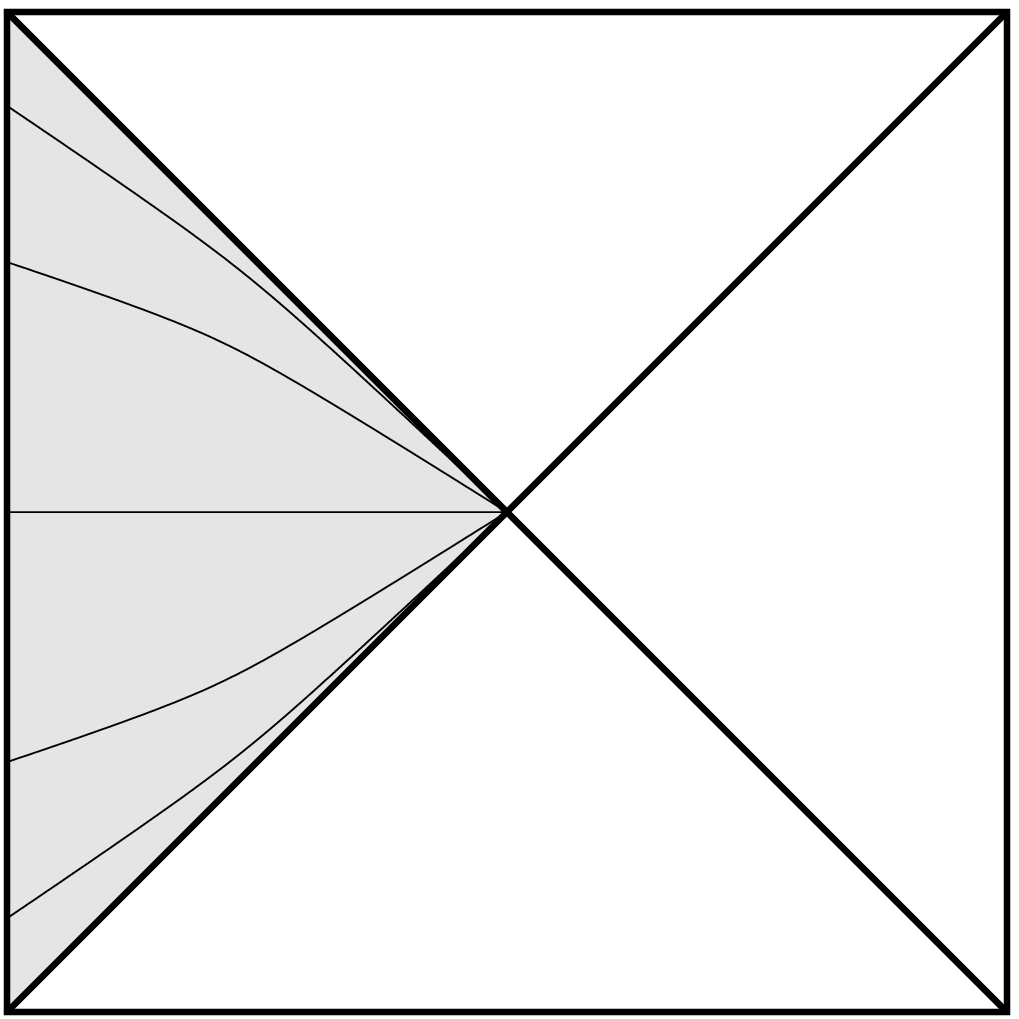,width=5cm}%
{Past and future event horizon (diagonal lines).  The static slicing
covers the interior of the cosmological horizon (shaded).%
\label{fig-static}}
In de~Sitter space, an observer is surrounded by a cosmological
horizon at $r=r_0$.  This is best seen in the static coordinate
system:
\begin{equation}
ds^2 = - V(r)\, dt^2 + \frac{1}{V(r)} dr^2 + r^2 d\Omega_2^2,
\end{equation}
where
\begin{equation}
V(r) = 1 - \frac{r^2}{r_0^2}.
\end{equation}
This system covers only part of the space-time, namely the interior of
a cavity bounded by $r=r_0$.  By the arguments given in
Sec.~\ref{sec-cd}, this is precisely the operationally meaningful
portion of de~Sitter space, because it is the largest causal diamond
possible.  It corresponds to a quarter of the Penrose diagram (e.g.,
for an observer at the left pole, the `left triangle' shown in
Fig.~\ref{fig-static}).

The upper and lower triangles contain exponentially large regions that
cannot be observed.  The spheres with $r=r_0$, the past and future
event horizons, are the entire diagonals of the square.  However, the
spheres with $r = r_0-\epsilon$ (the stretched
horizon~\cite{SusTho93}) lie within the left (or right) triangle and
represent the cosmological horizon of an observer at the corresponding
pole.

An object held at a fixed distance from the observer is redshifted;
the red-shift diverges near the horizon. \linebreak If released, the
object will accelerate towards the horizon.  If it crosses the
horizon, it cannot be retrieved.  Thus, the cosmological horizon acts
like a black hole `surrounding' the observer.  Note that the symmetry
of the space-time implies that the location of the cosmological
horizon is observer-dependent.

The black hole analogy carries over to the semi-classical
level~\cite{GibHaw77a}.  Because matter entropy can be lost when it
crosses, the cosmological horizon must be assigned a
Bekenstein-Hawking entropy equal to a quarter of its area, in order
for the generalized second law of
thermodynamics~\cite{Bek72,Bek73,Bek74} to remain valid:
\begin{equation}
S_{\rm 0} = \frac{A_0}{4},
\end{equation}
where
\begin{equation}
A_0 = 4 \pi r_0^2 = \frac{12 \pi}{\Lambda}
\end{equation}
The horizon emits Hawking radiation with temperature $(2\pi
r_0)^{-1}$.

\EPSFIGURE{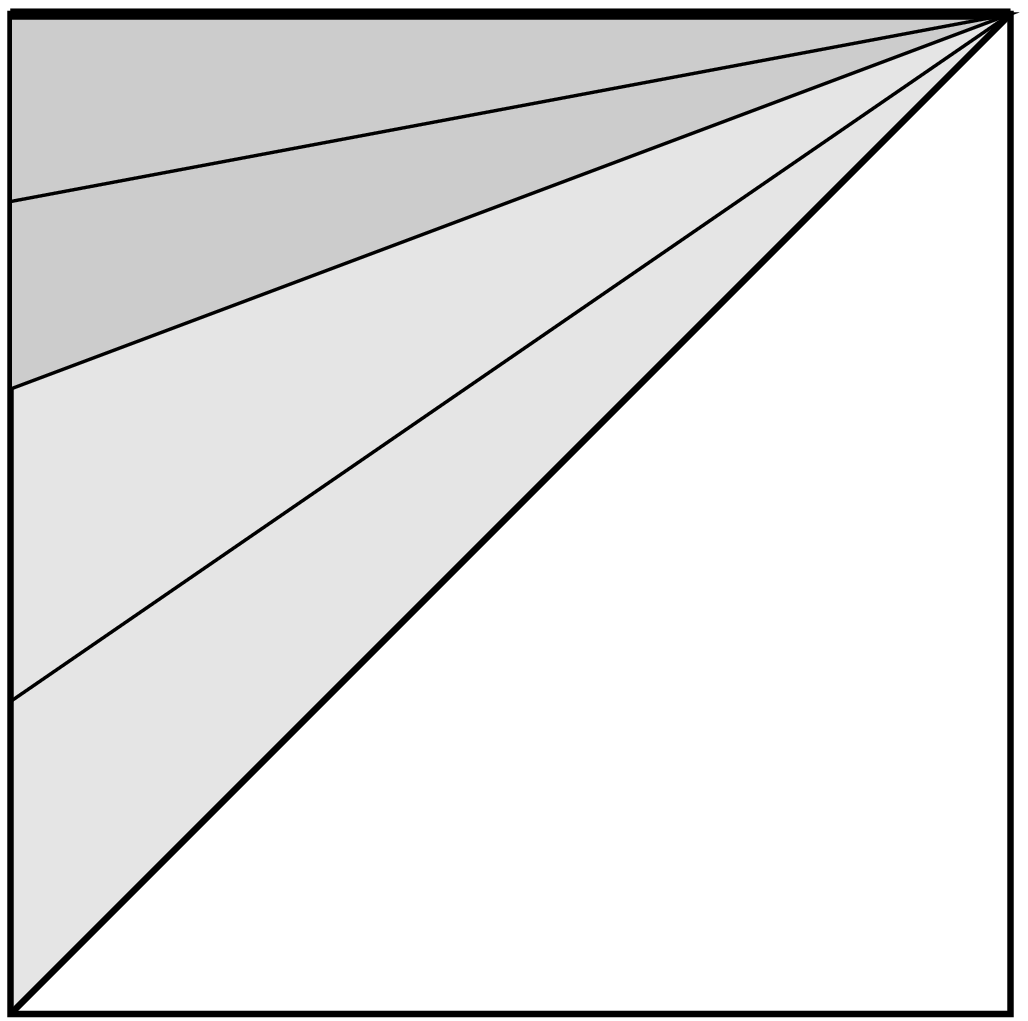,width=5cm}%
{The flat slicing covers half of de~Sitter space.  The dark shaded
region is the Penrose diagram of a flat big bang-de~Sitter cosmology
(Fig.~\ref{fig-R1}).
\label{fig-flat}}
de~Sitter space can also be written as a flat expanding FRW universe:
\begin{equation}
ds^2 = -d\tau^2 + \exp \left( \frac{2\tau}{r_0} \right) \left( dx^2 +
dy^2 + dz^2 \right).
\label{eq-flat}
\end{equation}
This metric covers half of the Penrose diagram (Fig.~\ref{fig-flat}).
If matter is present, it gives rise to a singularity on a space-like
slice at finite time $\tau_0$, which one can take to be $0$.  One thus
obtains a space-time which starts with a big bang and becomes
asymptotically de~Sitter in the future.  Its Penrose diagram is given
by a portion of the flat slicing, between some finite $\tau$ and
asymptotic infinity.

The remaining half of de~Sitter space is covered by the contracting
flat FRW universe obtained by time-reversal of Eq.~(\ref{eq-flat}).
By analogy with the previous paragraph, the introduction of matter
leads to a flat FRW universe that is asymptotically de~Sitter in the
past and collapses in a big crunch, with a time-reversed Penrose
diagram.  (These space-times are used in Sec.~\ref{sec-cd} to
illustrate the restriction to causal diamonds.)

\mbox{}\\[15ex]


\bibliographystyle{board}
\bibliography{all}

\end{document}